\documentclass[
    ,final            
  ]
  {aipproc}

\layoutstyle{8x11double}

\newcommand{\etal}{et al.}

\def\rosat{{\it ROSAT}}
\newcommand\chandra{{\it Chandra}}

\newcommand\xmm{{\it XMM-Newton}}

\def\ksnr{Kes~79}
\def\kpsr{{PSR J1852$+$0040}}
\def\esnr{{PKS~1209$-$51/52}}
\def\esrc{{1E~1207.4$-$5209}}
\def\epsr{{PSR~J1210$-$5226}}

\def\simlt{\mathrel{\hbox{\rlap{\hbox{\lower4pt\hbox{$\sim$}}}\hbox{$<$}}}}
\def\simgt{\mathrel{\hbox{\rlap{\hbox{\lower4pt\hbox{$\sim$}}}\hbox{$>$}}}}

\begin{document}

\title{CCO Pulsars as Anti-Magnetars: \\ Evidence of Neutron Stars Weakly Magnetized at Birth}


\author{E. V. Gotthelf}
{address={Columbia Astrophysics Laboratory, Columbia University,
New York, NY 10027}}
\author{J. P. Halpern}
{address={Columbia Astrophysics Laboratory, Columbia University,
New York, NY 10027}}
\begin{abstract}
Our new study of the two central compact object pulsars,
\epsr\ ($P = 424$~ms)
and PSR J1852$+$0040 ($P = 105$~ms), leads us
to conclude that a weak natal magnetic field shaped
their unique observational properties.  In the dipole spin-down
formalism, the 2-sigma upper limits on their period derivatives,
$< 2 \times 10^{-16}$ for both pulsars, implies surface magnetic
field strengths of $B_s < 3 \times 10^{11}$~G and spin periods at birth
equal to their present periods to three significant digits.  Their X-ray
luminosities exceed their respective spin-down luminosities, implying
that their thermal spectra are derived from residual cooling and perhaps
partly from accretion of supernova debris. For sufficiently weak
magnetic fields an accretion disk can penetrate the light cylinder
and interact with the magnetosphere while resulting torques on the
neutron star remain within the observed limits. We propose the
following as the origin of radio-quiet CCOs: the magnetic field,
derived from a turbulent dynamo, is weaker if the NS is formed
spinning slowly, which enables it to accrete SN debris. Accretion
excludes neutron stars born with both $B_s < 10^{11}$~G and $P > 0.1$~s
from radio pulsar surveys, where such weak fields are not encountered
except among very old ($> 40$~Myr) or recycled pulsars. We predict
that these birth properties are common, and may be attributes
of the youngest detected neutron star, the CCO in Cassiopeia~A, as well
as an undetected infant neutron star in the SN~1987A remnant.
In view of the far-infrared light echo discovered around Cas~A
and attributed to an SGR-like outburst, it is especially important
to determine via timing whether Cas~A hosts
a magnetar or not.  If not a magnetar, the Cas~A
NS may instead have undergone a one-time phase transition
(corequake) that powered the light echo.
\end{abstract}


\maketitle


\section{Introduction to CCO Pulsars}

The discovery in recent years of many isolated neutron stars (NSs) at
the centers of supernova remnants (SNRs) confirms the long-held notion
that these ultra-dense stellar remnants are born in supernova
explosions \cite{baa34}.  Most NSs are identified as
pulsars, whose emission derives either from rotational energy loss, as
for the rapidly spinning pulsars in the Crab ($P = 33$~ms) and Vela
($P = 89$~ms) remnants, or from magnetic field decay, as posited for
the highly magnetized $(10^{14} \simlt B_s \simlt 10^{15}$~G)
anomalous X-ray pulsars (AXPs) and 
soft gamma-ray repeaters (SGRs).  However, the
nature of a significant fraction of the young ($\simlt 10^4$~yrs)
NSs in SNRs is uncertain. These so-called central compact objects
(CCOs) are seemingly isolated NSs, distinguished by their steady flux,
predominantly thermal X-ray emission, lack of optical or radio counterparts,
and absence of a pulsar wind nebula (see review by
Pavlov \etal\ \cite{pav04}).

The properties of eight confirmed and proposed CCOs are summarized in Table~1.
(We omit the unique source 1E 161348$-$5055 at the center of
RCW 103 because its large-amplitude variability \cite{got99,del06}
violates the adopted definition.)
CCO luminosities are typically $10^{33}$
erg s$^{-1}$, similar to the younger pulsars; however, their spectra
are best characterized as hot blackbody emission of $kT_{BB} \sim
0.4$~keV, or two such components.  This is significantly
hotter than the surfaces of radio pulsars or other radio-quiet NSs
and corresponds to only a small fraction of the NS surface
area, with $R_{BB} \sim 0.7$~km.

Here we summarize new results on two unique pulsars that were initially
identified as CCOs.  Their properties, listed in Table 2, are
strikingly different from both the traditional radio pulsars
and the magnetars.  The lack of a measurable period derivative
requires that they have:
\begin{itemize}
\item small natal magnetic fields ($\simlt 10^{11}$~G),
\item slow initial spin periods very near to their present values
($>0.1$~s), and
\item X-ray luminosities powered by internal cooling, and possibly by accretion
of supernova (SN) debris.
\end{itemize}
\noindent
We predict that these properties will come to
characterize the CCO class of young neutron stars in general.
This can be tested with more sensitive searches for pulsations
from those objects listed in Table~1.

\begin{table}
\begin{tabular}{llcccrrl}
\hline\hline
{\hfil CCO \hfil} & {\hfil SNR \hfil} &
{Age} & {$d$}   & {$P$}  & {\hfil $f_p$\tablenote{Upper limits on pulsed fraction
are for a search down to $P=12$~ms or smaller.}\hfil }
& {\hfil $L_x$ \hfil} & {\hfil References \hfil} \cr
{   } & {   } & {(kyr)} & {(kpc)} & {(ms)} & (\%) & {(erg s$^{-1}$)} & \\
\hline
RX~J0822.0$-$4300 & Puppis~A           & 3.7  &  2   & \dots\   & $<5$ & $5 \times 10^{33}$ & \cite{hui06}\cr
CXOU~J085201.4$-$461753 & G266.1$-$1.2    & 1    &  1  & \dots\   & $<13$    & $2.5\times 10^{32}$ & \cite{sla01,kar02,bam05,iyu05}\cr
1E 1207.4$-$5209   & PKS~1209$-$51/52& 7   &  2 & 424      & 9         & $2.1 \times 10^{33}$   & \cite{zav00,mer02a,big03,del04}\cr
CXOU~J160103.1$-$513353 & G330.2$+$1.0     & $\simgt 3$ & 5 & \dots\ & \dots\ & $1.0 \times 10^{33}$ & \cite{par06}\cr
1WGA~J1713.4$-$3949 & G347.3$-$0.5    & 1.6   &  1.3    & \dots\   & $< 25$     & $1\times 10^{33}$ & \cite{laz03,cas04} \cr
CXOU~J181852.0$-$150213 & G15.9$+$0.2       & $1-3$ & (8.5) & \dots\   &  \dots\   & $1 \times 10^{33}$ & \cite{rey06}\cr
CXOU~J185238.6$+$004020 & Kes~79          & $7$  & 7   & 105      & 80        & $3.3 \times 10^{33}$ & \cite{sew03,got05,hal07}\cr
CXOU~J232327.9$+$584842 & Cas~A           & 0.33  & 3.4  & \dots\   & $<27$    & $2.6\times 10^{33}$ & \cite{pav00,cha01,mur02,mer02b}\cr
\hline
\end{tabular}\\
\caption{Central Compact Objects in Supernova Remnants}
\end{table}

\subsection{\epsr\ in SNR \esnr}

The central source \esrc\ in the SNR \esnr\ (Fig.~1)
is the most intensively
studied of the CCOs.  It acquired special importance as 
the first CCO detected as a pulsar \cite{zav00,pav02}.
It was distinguished again as the first isolated NS to display strong
absorption lines in its X-ray spectrum \cite{san02,mer02a,big03}.
We have made a comprehensive
study of all X-ray timing data on \esrc\ (\epsr) \cite{got07},
showing that its rotation is highly stable (Fig. 2), with $P =
424.130749(4)$~ms and $\dot P = (6.6 \pm 9.0) \times 10^{-17}$
($1\sigma$ errors), superseding previous claims of large period
changes in the same data \cite{zav04,woo07}. In
the dipole spin-down formalism, the $2\sigma$ upper limit on $\dot P$
implies a spin-down luminosity $\dot E \equiv -I\Omega\dot \Omega < 1.3 \times
10^{32}$~erg~s$^{-1}$, surface magnetic field strength $B_s < 3.3
\times 10^{11}$~G, and characteristic age $\tau_c \equiv P/2\dot P>
27$~Myr.  This lower limit on $\tau_c$ exceeds the SNR age by at least
$3$ orders of magnitude, requiring that the pulsar was born spinning
at its present period to three significant digits.  The X-ray
luminosity of \esrc, $L_{\rm bol} \approx 2.1 \times 10^{33}\,(d/2\ {\rm
kpc})^2$ erg~s$^{-1}$, exceeds its $\dot E$, implying that $L_{\rm
bol}$ derives from residual cooling, and perhaps partly from accreting
SN debris.

\subsection{\kpsr\ in SNR Kes 79}

We also discovered pulsations from a second CCO, in the
SNR Kes~79 \cite{got05} (Fig.~3).
Our follow-up program to time
\kpsr\ produced a remarkable result: no change in its 105~ms
period over 2.4 yr \cite{hal07}.  From the data
shown in Figure~4, we derived a
$2\sigma$ upper limit of $\dot P < 2.0 \times 10^{-16}$, which
leads to a spin-down luminosity $\dot E < 7 \times 10^{33}$~erg~s$^{-1}$,
surface magnetic field strength $B_s < 1.5 \times 10^{11}$~G,
and characteristic age $\tau_c > 8$~Myr.
Again, this implies that the pulsar was born spinning
at its current period, with
a weaker $B$-field than that of any other young pulsar.

These are the only two NSs whose initial rotation periods are so precisely
inferred.  They are longer than what was once thought typical,
but in fact are consistent with recent statistical
analyses of the radio pulsar population, e.g.,
Faucher-Gigu\`ere \& Kaspi \cite{fau06} who
favor a wide distribution of birth periods
(Gaussian mean $P \sim 300$~ms and $\sigma_P \sim 150$~ms).
The absence of spin-down for both CCO pulsars then
highlights the difficulty of existing theories to
explain the high luminosities and temperatures of CCO thermal X-ray
spectra. Their
X-ray luminosities are a large fraction of their $\dot E$, challenging the
rotation-powered assumption, and greater than their reservoirs of
$B$-field energy, refuting the magnetar hypothesis.

\begin{table}
\begin{tabular}{lcc}
\hline\hline
{\hfil Parameter \hfil} & {PSR J1209$-$5226} & {PSR J1852$+$0040} \cr
\hline
$P$ (ms)                 & 424.1307 & 104.9126 \cr
$\dot P$ ($10^{-17}$ s s$^{-1}$)   & $6.6 \pm 9.0$ & $-34 \pm 27$  \cr
$\dot E$ ($10^{32}$ erg s$^{-1}$)\tablenote{$2\sigma$ limit assuming
magnetic dipole spin-down.}
& $<1.3$ & $<70$ \cr
$L({\rm bol})/\dot E$
& $> 15$ & $>0.5$ \cr
$B_s$ ($10^{11}$ G)$^*$              & $<3.3$   & $<1.5$   \cr
$\tau_c$ (Myr)$^*$ & $>27$   & $>8$     \cr
SNR age (kyr)                       & $\sim 7$ & $\sim 7$ \cr
Reference & \cite{got07} & \cite{hal07} \cr
\hline
\end{tabular}\\
\caption{Spin Parameters of CCO Pulsars}
\end{table}

\section{Accreting and/or cooling}

Instead, the high blackbody temperatures ($kT_{BB} \sim 0.4$~keV)
and small blackbody radii ($R_{BB} \sim 0.7$~km) of CCOs
may be evidence of accretion onto a polar cap, possibly
from a fallback disk of SN debris.  Interior cooling models, even
with anisotropic conduction, do not make such a concentrated hot spot.
We \cite{hal07} proposed this as the origin of CCOs: 
Magnetic field is generated by a turbulent dynamo
whose strength depends on the rotation rate of the
proto-neutron star \cite{tho93,bon06},
so the magnetic field strength would be inversely correlated with
the initial period.  If $P$ is large and $B_s$ is small,
accretion of SN debris is possible.   Accretion excludes
neutron stars born with both $B_s < 10^{11}$~G 
and $P > 0.1$~s from radio pulsar surveys,
where $B_s < 10^{11}$~G is not found except in very old
($\tau_c > 40$~Myr) or recycled pulsars.

\begin{figure}
\includegraphics[height=0.32\textheight]{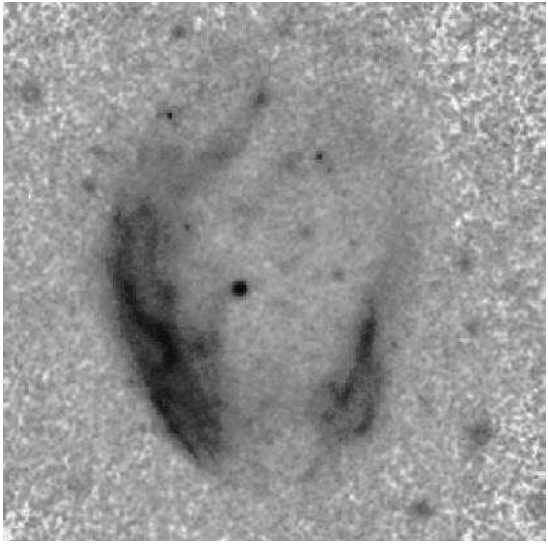}
\caption{
Greyscale \rosat\ X-ray image of the CCO pul-\break sar \epsr\ 
in the SNR \esnr.  The pulsar is offset from the center of the
barrel-shaped thermal remnant.}
\end{figure}

\begin{figure}
\includegraphics[width=0.33\textheight]{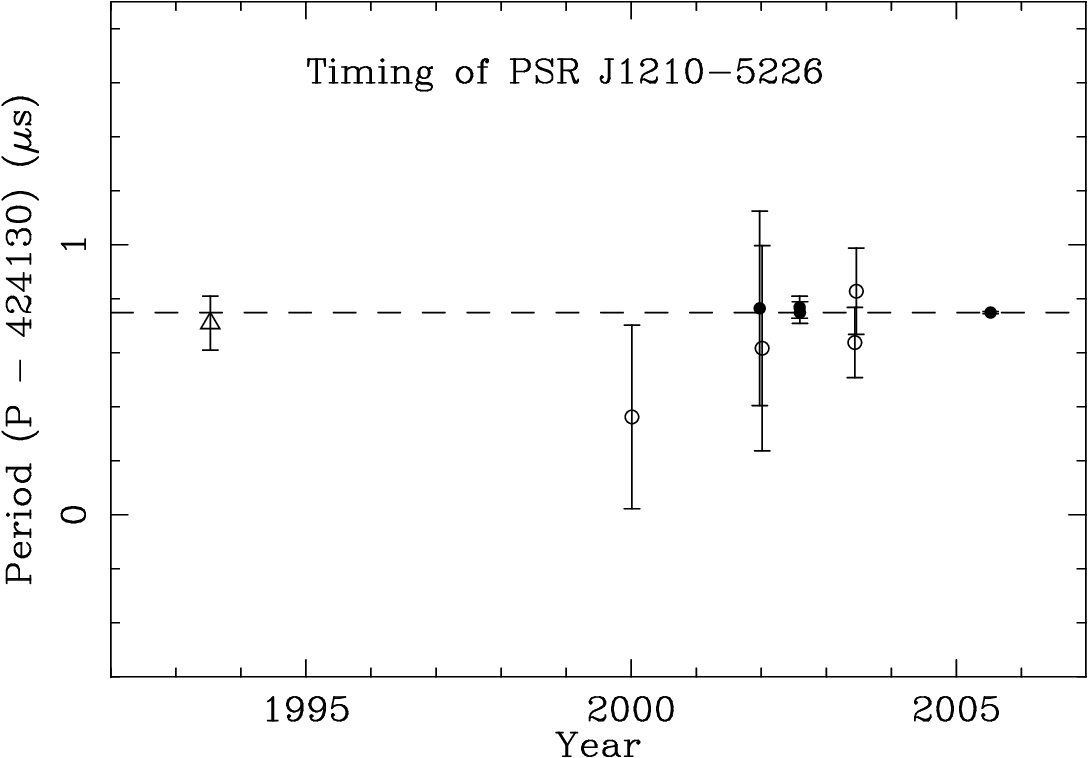}
\caption{Timing of \epsr\ \cite{got07}.
{\it Filled circles} are from \xmm, {\it open circles} are from \chandra,
and the {\it open triangle} is from \rosat.}
\end{figure}

Another clue to the nature of CCOs is the soft X-ray
spectrum of \esrc.  It has broad absorption lines
centered at $0.7$~keV and $1.4$~keV \cite{san02,mer02a,big03,del04}.
Our upper limit, $B_s < 3.3 \times 10^{11}$~G, favors the electron
cyclotron model \cite{big03,del04},
for at least one of the lines, over all others that
require stronger fields.  The basic cyclotron prediction, $B_s = 8 \times
10^{10}$~G, assumes that 0.7~keV is the fundamental energy $E_c =
1.16(B_s/10^{11}\,{\rm G})/(1+z)$ keV, where $z$ is the gravitational
redshift. Another solution postulates hydrogenic oxygen for the
0.7~keV line, while the 1.4~keV line is the cyclotron fundamental
\cite{hai02,mor06}.  As the authors of the latter hypothesis pointed out,
abundant oxygen may be accreted from SN debris.

\begin{figure}
\includegraphics[height=0.32\textheight]{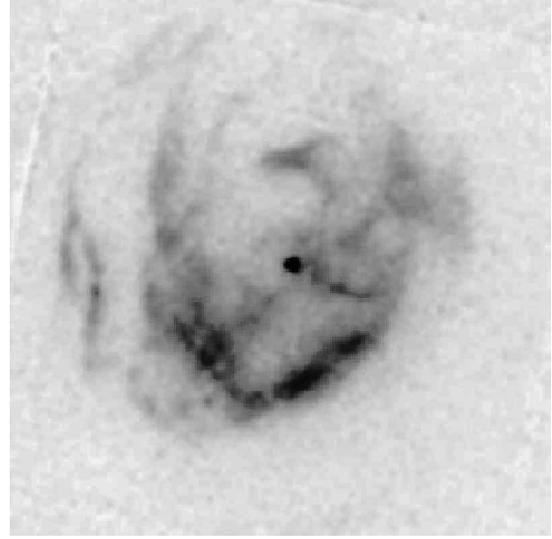}
\caption{
Greyscale \xmm\ X-ray image of the CCO pulsar \kpsr\ in the SNR \ksnr.
The image is centered on the pulsar within the shell-type thermal remnant.}
\end{figure}

\begin{figure}
\includegraphics[width=0.33\textheight]{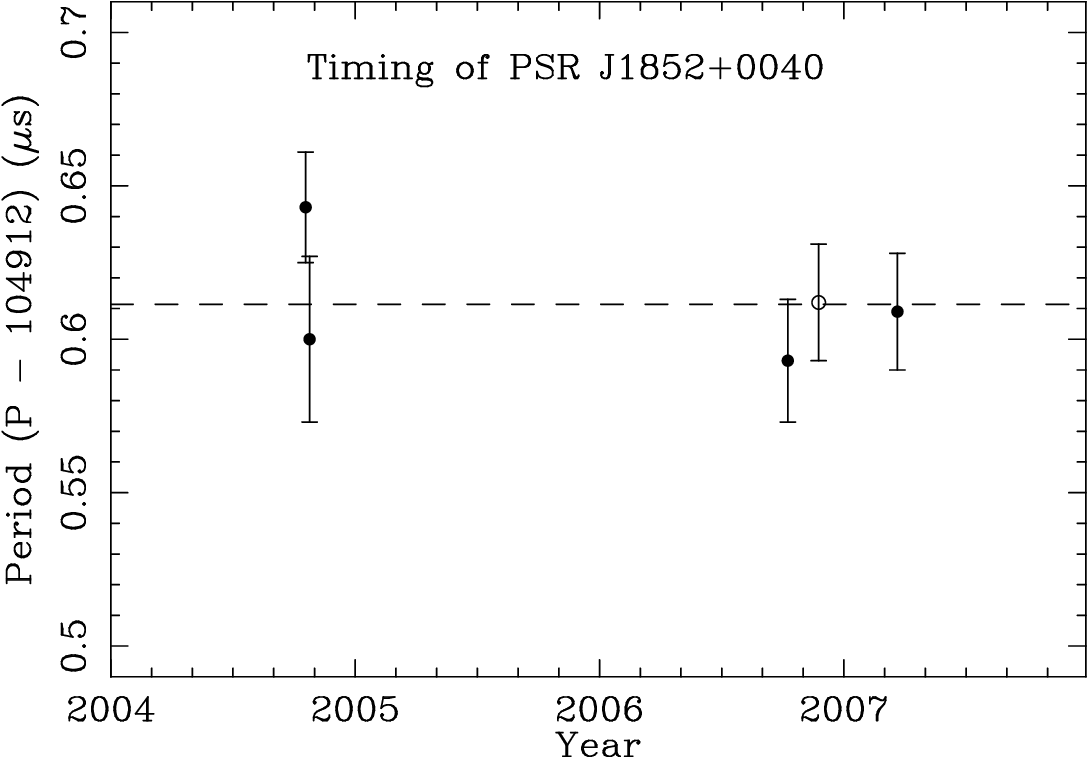}
\caption{ Timing of \kpsr\ \cite{hal07}. 
{\it Filled circles} are from \xmm\ and the {\it open circle} 
is from \chandra.}
\end{figure}

Accretion from a fallback disk of SN debris in the propeller
regime has been considered for CCOs by several authors
\cite{alp01,kar02,shi03,zav04,eks05,liu06}.
We propose that this may be the first
phase in the life of those neutron stars born rotating slowly
with weak magnetic fields.
The X-ray luminosity of CCOs, or possibly just their hot spots,
can be powered
by accretion of $\dot m \approx 3 \times 10^{13}$ g~s$^{-1}$,
or only $\approx 0.1$ lunar masses of supernova debris over
their $\sim 7$~kyr lifetimes.
The main barrier to disk accretion is the speed-of-light cylinder,
of radius $r_{\ell} = cP/2\pi$.
If an accretion disk cannot penetrate the light cylinder, the NS
cannot interact with the disk, and it behaves as a isolated pulsar.
But if $B_s$ is as small as $10^{10}$~G, accretion at a rate
$\dot M \geq 10^{13}$~g~s$^{-1}$ can penetrate the light cylinder
to the magnetospheric radius $r_m$, since $r_m < r_{\ell}$
in this case.
If so, the system is in the propeller regime,
in which matter flung out from
$r_m$ at a rate $\dot M$ takes angular momentum from the NS,
causing it to spin down.
In the case of \kpsr, we \cite{hal07}
estimated the propeller spin-down rate as
\begin{displaymath}
\dot P \approx 2.2 \times 10^{-16}\,\mu_{28}^{8/7}\,\dot M_{13}^{3/7}
\left({M \over M_{\odot}}\right)^{-2/7}
\end{displaymath}
\begin{equation}
\times \ I_{45}^{-1}\left({P \over 0.105\ {\rm s}}\right)
\left(1- {P \over P_{\rm eq}}\right)
\end{equation}
using the prescription of Menou \etal\ \cite{men99}.
Here $I \approx 10^{45}$ g~cm$^2$
is the NS moment of inertia,
$\mu = B_s\,R^3 \approx 10^{28}$ G~cm$^3$,
and $P_{\rm eq}$ is the equilibrium, or minimum
period for disk accretion.
Appropriate scaling for the 0.424~s period of \esrc\ can be
substituted in equation (1).  These predictions for accretion are close to
the observed upper limits on $\dot P$ (Table~2).

We distinguish here between $\dot M$, the matter expelled that is
responsible for the torque on the NS,
and $\dot m \approx 3 \times 10^{13}$~g~s$^{-1}$,
the matter accreted, which is responsible for the X-ray emission
from the surface, presumably at a magnetic pole of the NS.
For the propeller model to be self-consistent,
$\dot M$ must be  $> \dot m$, which is possible according to
equation (1) as long as $B_s < 10^{10}$~G.
In X-ray binaries, the propeller effect does not necessarily
preclude accretion onto the NS surface.
The disposition of material leaving the inner
edge of the accretion disk is not well understood,
and many authors, e.g., Rappaport \etal\ \cite{rap04}, consider that
accretion onto the NS can proceed even in the propeller (fast pulsar)
regime.  We adopt that point of view here.
Accreting X-ray binary pulsars are often found in near equilibrium
states where $\dot P$ changes sign without changing luminosity.
Equation (1) may then represent the typical magnitude of $\dot P$,
independent of sign, albeit at an accretion rate 4--5 orders of magnitude
less than in binaries.

Whether CCO pulsars are accreting or not,
the conclusion that they have weak magnetic fields is
unavoidable.  The hypothesis of dipole spin-down can be
quantified by measuring $\dot P$ using phase-coherent timing,
the only method that is effective in a reasonable time span
for detecting such small $B$-fields.  A steady,
positive $\dot P$ will yield $B_s$ via the dipole spin-down formula,
$B_s = 3.2 \times 10^{19}(P \dot P)^{1/2}$. But if the pulsars 
show episodes of fluctuating or negative $\dot P$, that
will be evidence of accretion, i.e., torque noise.

Existing data cannot (yet) be used to accomplish these tests
because they were not densely spaced enough to span gaps of
several months with a coherent timing solution free of cycle
count uncertainties.  But a carefully planned sequence of
new observations can be used to fit a coherent
ephemeris that will bootstrap
the historical data into a phase-linked timing solution
spanning 4.5 years for \kpsr, and 10 years for \epsr,
which will improve the sensitivity to $\dot P$ by 3 orders of magnitude
over the limits in Table~2, reaching limits of $B_s \sim 10^{10}$~G.

\section{Are There Younger CCOs?}

CCOs represent a significant fraction of young NSs.
Here, we propose that their birth properties
are common enough to be a likely explanation
for the inconspicuous nature of the two youngest
NSs, the CCO in Cas~A, and an undetected pulsar
in the SN 1987A remnant.

\subsection{Cas A: Magnetar or Anti-Magnetar?}

The Cas~A CCO is the youngest
known NS (327 yr); the SN was probably
seen by Flamsteed in 1680, which agrees with the measured
convergent date of the SNR ejecta \cite{tho01}.
The simple argument that CCOs
are born spinning at their current periods, and with their
current magnetic field strengths, implies that their
luminosities need not decrease substantially since
birth.  Only their interior cooling would reduce
their soft X-ray emission. 
The Cas~A CCO is characterized, like the others,
by its steady X-ray luminosity of $\approx 2.6 \times 10^{33}$ erg~s$^{-1}$,
compared to $\sim 10^{35}$ erg~s$^{-1}$ for magnetars.

Nevertheless, despite the strong evidence that CCO pulsars
have weak magnetic fields, there is circumstantial
evidence that Cas A may host a magnetar.
Rapidly moving IR features detected by the
{\it Spitzer Space Telescope} outside Cas~A were interpreted as a light echo
from a beamed, high-energy flare some 55 years ago that is heating the
surrounding interstellar dust \cite{kra05}.
An SGR-like outburst of $\sim 2 \times 10^{46}$~erg (isotropic equivalent)
would have been required to explain the reprocessed IR luminosity.
Until recently, a quiescent magnetar was a favored
hypothesis for the nature of the Cas~A point source.

However, the magnetar hypothesis implies that the present
rotation period of the 327~old pulsar is
$P \approx 0.45(B_s/10^{14}\ {\rm G})$~s,
assuming that its initial period $P_0<<P$,  which also requires
that $\dot E \approx 9.5 \times 10^{36}(B_s/10^{14}\ {\rm G})^{-2}$
erg~s$^{-1}$. Cas~A could therefore have a spin-down
luminosity that exceeds the typical $10^{35}$ erg~s$^{-1}$ X-ray
luminosity of all other magnetars.
Such a large spin-down luminosity is almost always
accompanied by substantial non-thermal X-ray luminosity, of order
$10^{-3} \dot E$, including a resolvable
pulsar wind nebula.  The absence of any X-ray evidence of
such energetic spin-down argues against the presence of
a typical magnetar $B$-field in the Cas~A point source.

Thus, Cas~A is the
focus of sharply contradictory evidence about the birth properties
of CCOs: their natal magnetic
field strengths and initial spin periods.
Apart from
waiting for an SGR or AXP-like outburst (which may never occur),
the only way to resolve whether Cas~A hosts
a transient magnetar or an ``anti-magnetar''
is by direct measurement of its spin properties.
A pulsar in Cas~A will reveal the spin period and dipole
$B$-field of a NS at an age that is only a few percent of the
known AXPs' and CCOs' ages,
providing a correspondingly more secure representation
of their initial values.  If the magnetar hypothesis is not confirmed,
it may imply that the high-energy flare was powered by a one-time event
in Cas A, a first-order phase transition (corequake)
\cite{zdu07}, e.g., to a pion or kaon condensate or deconfined quarks.
That would be a remarkable outcome.

\subsection{\bf Is there a CCO in SN~1987A?}

It has long been known that the non-detection of a pulsar in SN 1987A
can be explained if the NS was born with a weak $B$-field or a long
rotation period \cite{oge04,man07}. Now, we have established that a
CCO can have both, which gives it essentially the same spin-down
luminosity at birth that it has at an age of $10^4$~yr.  This means
that a CCO can emit less than the observed limits from SN~1987A even
if 100\% of its spin-down power is reprocessed into IR emission by
dust in the surrounding SN ejecta.  The observed luminosity limits for
a point source inside the ring of SN~1987A are -- in X-rays:
$L_x(2-10\,{\rm keV}) < 1.5 \times 10^{34}$ erg~s$^{-1}$ corrected for
extinction \citep{par04}, in the visible range: $L(2900-9650$~\AA) $<
8 \times 10^{33}$ erg~s$^{-1}$ corrected for dust absorption
\cite{gra05}, and at mid-IR wavelengths: $L(10\,\mu{\rm m}) < 1.5
\times 10^{36}$ erg~s$^{-1}$ for dust emitting at $T \approx 90-100$~K
\cite{bou04}.  This mid-IR luminosity can be accounted for by
radioactive decay of $^{44}$Ti, and therefore represents a very
conservative upper limit on the spin-down power of an embedded pulsar.
At an age of $10-20$~yr, a cooling NS need emit only $\approx 3 \times
10^{34}$ erg~s$^{-1}$ of soft X-rays at a temperature of $2.5 \times
10^6$~K \cite{yak02}, some of which is absorbed by SN ejecta or ISM in
the LMC.  So we conclude that a CCO is a promising model for an unseen
NS in SN~1987A.  Given the lack of any other NS signatures from CCOs,
a continuing search for surface thermal X-ray emission as the ejecta
thin out is perhaps the best hope of detecting the NS in SN~1987A,
even though it is becoming more difficult as the SNR brightens
dramatically.


\begin{theacknowledgments}
This work uses data obtained with the  \xmm\ and
\chandra\ observatories and funded by NASA grants NNX06AH95G and SAO
G06-7048X. EVG thanks the conference organizers for financial support.
\end{theacknowledgments}


\bibliographystyle{aipprocl} 



\begin{thebibliography}{9}
\bibitem{baa34}Baade, W., \& Zwicky, F., Phys. Rev. 45, 138 (1934).
\bibitem{pav04}Pavlov, G. G., Sanwal, D., \& Teter, M. A.,
IAU Symp. 218, 239 (2004).
\bibitem{hui06}Hui, C. Y., \& Becker, W., A\&A 454, 543 (2006).
\bibitem{sla01}Slane, P., \etal, ApJ 548, 814 (2001).
\bibitem{kar02}Kargaltsev, O., \etal, ApJ 580, 1060 (2002).
\bibitem{bam05}Bamba, A., Yamazaki, R., \& Hiraga, J. S., ApJ 632, 294 (2005).
\bibitem{iyu05}Iyudin, A. F., \etal, A\&A 429, 225 (2005).
\bibitem{zav00}Zavlin, V. E., \etal, ApJ 540, L25 (2000).
\bibitem{mer02a}Mereghetti, S., \etal, ApJ 581, 1280 (2002).
\bibitem{big03}Bignami, G. F., \etal, Nature 423, 725 (2003).
\bibitem{del04}De Luca, A., \etal, A\&A 418, 625 (2004).
\bibitem{par06}Park, S., \etal, ApJ 653, L37 (2006).
\bibitem{laz03}Lazendic, J. S., \etal, ApJ 593, L27 (2003).
\bibitem{cas04}Cassam-Chena\"i, G., \etal, A\&A 427, 199 (2004).
\bibitem{rey06}Reynolds, S. P., \etal, ApJ 652, L45 (2006).
\bibitem{sew03}Seward, F. D., \etal, ApJ 584, 414 (2003).
\bibitem{got05}Gotthelf, E. V., Halpern, J. P., \& Seward, F. D.,
ApJ 627, 390 (2005).
\bibitem{hal07}Halpern, J. P., \etal, ApJ 665, 1304 (2007).
\bibitem{pav00}Pavlov, G. G., \etal, ApJ 531, L53 (2000).
\bibitem{cha01}Chakrabarty, D., \etal, ApJ 548, 800 (2001).
\bibitem{mur02}Murray, S. S., \etal, ApJ 566, 1039 (2002).
\bibitem{mer02b}Mereghetti, S., Tiengo, A., \& Israel, G. L.,
ApJ 569, 275 (2002).
\bibitem{got99}Gotthelf, E. V., Petre, R., \& Vasisht, G.,
ApJ 514, L107 (1999).
\bibitem{del06}De Luca, A., \etal, Science 313, 814 (2006).
\bibitem{got07}Gotthelf, E. V., \& Halpern, J. P., ApJ 664, L35 (2007).
\bibitem{pav02}Pavlov, G., G., \etal, ApJ 569, L95 (2002).
\bibitem{san02}Sanwal, D., \etal, ApJ 574, 61 (2002).
\bibitem{zav04}Zavlin, V. E., Pavlov, G. G., \& Sanwal, D.,
ApJ 606, 444 (2004).
\bibitem{woo07}Woods, P. M., Zavlin, V. E., \& Pavlov, G. G.,
Ap\&SS 308, 239 (2007).
\bibitem{fau06}Faucher-Gigu\`ere, C.-A., \& Kaspi, V. M., ApJ 643, 332 (2006).
\bibitem{tho93}Thompson, C., \& Duncan, R. C., ApJ 408, 194 (1993).
\bibitem{bon06}Bonanno, A., Urpin, V., \& Belvedere, G., A\&A 451, 1049 (2006).
\bibitem{hai02}Hailey, C. J., \& Mori, K., ApJ 578, L133 (2002).
\bibitem{mor06}Mori, K., \& Hailey, C. J., ApJ 648, 1139 (2006).
\bibitem{alp01}Alpar, M. A., ApJ 554, 1245 (2001).
\bibitem{shi03}Shi, Y., \& Xu, R. X., ApJ 596, L75 (2003).
\bibitem{eks05}Ek\c si, K. Y., \etal, ApJ 623, L41 (2005).
\bibitem{liu06}Liu, D. B., \etal, ApJ 644, 439 (2006).
\bibitem{men99}Menou, K., \etal, ApJ 520, 276  (1999).
\bibitem{rap04}Rappaport, S. A., Fregeau, J. M., \& Spruit, H., ApJ 606,
436 (2004).
\bibitem{tho01}Thorstensen, J. R., Fesen, R. A., \& van den Bergh, S., 
AJ 122, 297 (2001).
\bibitem{kra05}Krause, O., \etal, Science 308, 1604 (2005).
\bibitem{zdu07}Zdunik, J. L., \etal, A\&A 465, 533 (2007).
\bibitem{oge04}\"Ogelman, H., \& Alpar, M. A., ApJ 603, L33 (2004).
\bibitem{man07}Manchester, R. N., AIP Conf. Proc. 937, 134 (2007).
\bibitem{par04}Park, S., \etal, ApJ 610, 275 (2004).
\bibitem{gra05}Graves, G. J. M., \etal, ApJ 629, 944 (2005).
\bibitem{bou04}Bouchet, P., \etal, ApJ 611, 394 (2004).
\bibitem{yak02}Yakovlev, D. G., \etal, A\&A 389, L24 (2002).
\end{thebibliography}


\end{document}